\documentclass{article}
\usepackage{latexsym,amssymb,amsfonts,amsmath}
\usepackage[dvipdfmx]{graphicx}

\newtheorem{thm}{Theorem}[section]
\newtheorem{lem}[thm]{Lemma}
\newtheorem{cor}[thm]{Corollary}
\newtheorem{pro}[thm]{Proposition}

\newtheorem{conj}[thm]{Conjecture}

\newcommand{\RM}{\mathbb{R}}
\newcommand{\ZM}{\mathbb{Z}}
\newcommand{\ZMP}{\mathbb{Z}_{\geq}}

\newcommand{\CM}{\mathbb{C}}

\newcommand{\DM}{\mathbb{D}}

\newcommand{\ket}[1]{|#1\rangle}
\newcommand{\bra}[1]{\langle#1|}

\begin{document}
\setlength{\textheight}{8.0truein}    

\title{RETURN PROBABILITY AND SELF-SIMILARITY OF THE RIESZ WALK}

\author{RYOTA HANAOKA\thanks{Department of Applied Mathematics, Graduate School of Engineering Science, Yokohama National University, Tokiwadai, Hodogaya, Yokohama, 240-8501, Japan} \and NORIO KONNO\thanks{Department of Applied Mathematics, Graduate School of Engineering, Yokohama National University, Tokiwadai, Hodogaya, Yokohama, 240-8501, Japan}}

\date{}

\maketitle
\begin{abstract}
	The quantum walk is a counterpart of the random walk. The 2-state quantum walk in one dimension can be determined by a measure on the unit circle in the complex plane. As for the singular continuous measure, results on the corresponding quantum walk are limited. In this situation, we focus on a quantum walk, called the Riesz walk, given by the Riesz measure which is one of the famous singular continuous measures. The present paper is devoted to the return probability of the Riesz walk. Furthermore, we present some conjectures on the self-similarity of the walk.
\end{abstract}

{\small {\it keywords}: Quantum walks, Singular continuous measure, Return probability, Self-similarity}

\vspace*{1pt}    
\section{Introduction}        
Quantum walk (QW) was introduced as a quantum version of random walk (RW) and has been widely studied since around 2000 \cite{AharonovEtAl1993,AmbainisEtAl2001,Gudder1988,Meyer1996}. Some properties that appear in QWs but not in RWs are linear spreading and localization. One of the approaches to the study on QWs is the CGMV method introduced by Cantero, Gr\"unbaum, Moral, and Vel{\'a}zquez \cite{Cantero2012}. By using this method, we can associate a QW with a measure on the unit circle in the complex plane. This method has explained the characteristics of QWs, such as recursion and localization conditions \cite{Bourgain2014,Grunbaum2013}. The QW with a singular measure on the unit circle, considered in this paper, has been studied by some researchers. For example, it was shown in \cite{Cedzich2020} that a QW with the coin of each location determined by irrational number in a magnetic field has its spectrum as the Cantor set. There are other related results on singular measures and the CMV matrices, such as quantum intermittency \cite{DamanikEricksonEtAl2016}, Simon's subshift conjecture \cite{DamanikLenz2007}, the singular continuous spectrum for OPUC \cite{Ong2012}, and the period doubling subshift \cite{Ong2014}. A typical example of the QW with a singular measure was introduced by Gr\"unbaum and Vel{\'a}zquez \cite{Grunbaum2012} in 2012, where measure is given by the Riesz one. In \cite{Khrushcev2001,Simon2005}, the Schur parameter is given by the Riesz product. However, the results of the Riesz measure are not much obtained. Following Gr\"unbaum and Vel{\'a}zquez \cite{Grunbaum2012}, we call the QW defined by the Riesz measure {\it the Riesz walk}.\par
In this paper, we compute the return probability of the Riesz walk. For the Riesz walk on a half line starting from the origin, we calculate the measure of the origin for any time. Therefore, we obtain a specific behavior corresponding to the singular continuous measure. Furthermore, numerical simulations suggest some interesting conjectures of the evolution of the Riese walk.\par
The rest of this paper is organized as follows. Section \ref{def} gives the definition of the Riesz walk. In Section \ref{returnsec}, we present our main result (Theorem \ref{returnthm}) related to the return probability at the origin. Section \ref{conjsec} shows some conjectures on the self-similarity of the Riesz walk suggested by numerical calculations. In Section \ref{SecGeneral}, we consider QWs consisting of singular continuous measures, which are related to Riesz walks. Section \ref{conclusion} summarizes this paper.

\section{Defintions}\label{def}
\subsection{CGMV method}\label{cgmv}
The QW on $\ZMP$, considered in this paper, can be determined by a measure $\mu$ on the unit circle $\partial\DM$ in the complex plane $\CM$, where $\ZMP=\{0,1,2,\ldots\}$, $\partial\DM=\{z\in\CM:|z|=1\}$, and $\CM$ is the set of complex numbers. Given a measure $\mu$ on $\partial\DM$, the Carath\'eodory and Schur functions are defined as follows.
\begin{align*}
	F(z)=\int\frac{t+z}{t-z}d\mu(t),\quad f(z)=\frac{1}{z}\frac{F(z)-1}{F(z)+1}.
\end{align*}
Then, the Verblunsky parameters $\alpha_k$ are given by the following algorithm.
\begin{align*}
	f_{k+1}(z)=\frac{1}{z}\frac{f_{k}(z)-f_{k}(0)}{1-\overline{f_{k}(0)}f_{k}(z)},\quad \alpha_{k}=f_{k}(0)\quad ( k\geq 0),
\end{align*}
where $f_{0}(z)=f(z)$. From the Verblunsky parameters, the corresponding QW on $\ZMP$ is determined by
\begin{align*}
	&U^{(s)}=\begin{bmatrix}
		R_0&Q_1&O&O&O&\ldots\\
		P_0&R_1&Q_2&O&O&\ldots\\
		O&P_1&R_2&Q_3&O&\ldots\\
		O&O&P_2&R_3&Q_4&\ldots\\
		O&O&O&P_3&R_4&\ldots\\
		\vdots&\vdots&\vdots&\vdots&\vdots&\ddots
	\end{bmatrix},\\
	&R_{0}=\begin{bmatrix}\overline{\alpha_{0}}&\rho_{0}\\\rho_{0}\overline{\alpha_{1}}&-\alpha_{0}\overline{\alpha_{1}}\end{bmatrix},\quad P_{x}=\begin{bmatrix}\rho_{2x}\rho_{2x+1}&-\alpha_{2x}\rho_{2x+1}\\0&0\end{bmatrix},\\ &R_{x}=\begin{bmatrix}-\alpha_{2x-1}\overline{\alpha_{2x}}&-\alpha_{2x-1}\rho_{2x}\\\rho_{2x}\overline{\alpha_{2x+1}}&-\alpha_{2x}\overline{\alpha_{2x+1}}\end{bmatrix},\quad Q_{x}=\begin{bmatrix}0&0\\\rho_{2x-1}\overline{\alpha_{2x}}&\rho_{2x-1}\rho_{2x}\end{bmatrix},
\end{align*}
where $\rho_k=\sqrt{1-|\alpha_k|^2}$ and $\overline{\alpha_k}$ is the complex conjugate of $\alpha_k$. Note that the transpose of $U^{(s)}$ is used in orthogonal polynomial literature in Simon \cite{Simon2005}. The amplitude of location $x\in\ZMP$ at time $t\in\ZMP$ is denoted by $\Psi_t(x)=[\Psi_t^{L}(x),\Psi_t^{R}(x)]^{\mathrm{T}}$, where $\mathrm{T}$ is the transposed operator. The evolution is defined by the equation $\Psi_{t+1}=U^{(s)}\Psi_t$, where $\Psi_t=[\Psi_t(0),\Psi_t(1),\Psi_t(2),\ldots]^{\mathrm{T}}$. Then the probability of location $x$ at time $t$ is given by $\mu_t(x)=\left\|\Psi_t(x)\right\|^2=\left|\Psi_t^{L}(x)\right|^2+\left|\Psi_t^{R}(x)\right|^2$.
\subsection{Riesz walk}
The Riesz measure on the unit circle $\partial\DM$ in $\CM$ is defined by
\begin{align*}
	d\mu(z)=\prod_{k=1}^{\infty}\left(1+\cos(4^k\theta)\right)\frac{d\theta}{2\pi}=\prod_{k=1}^{\infty}\left(1+\frac{z^{4^k}+z^{-4^k}}{2}\right)\frac{dz}{2\pi iz}=\sum_{j=-\infty}^{\infty}\overline{\mu}_{j}z^{j}\frac{dz}{2\pi iz},
\end{align*} 
where $z=e^{i\theta}$ with $\theta\in[0,2\pi)$ and $\mu_j$ is the $j$-th moment of $\mu$. The $\mu_j$ can be written as follows (see \cite{Grunbaum2012}).
\begin{align}\label{rieszmoment}
	\mu_{j}=\left\{ \begin{array}{cl}
		1,&j=0,\\
		1/2^{p},&j=\pm4^{k_1}\pm4^{k_2}\pm\cdots\pm4^{k_p},\\
		0,&otherwise,
	\end{array}\right. 
\end{align}
where $k_1>k_2>\cdots>k_p\geq1$. The Carath\'eodory and Schur functions of the Riesz measure are computed in the following fashion.
\begin{align*}
	F(z)&=1+2\sum^{\infty}_{n=1}\overline{\mu}_nz^n=1+z^{4}+\frac{z^{12}}{2}+z^{16}+\frac{z^{20}}{2}+\frac{z^{44}}{4}+\frac{z^{48}}{2}+\frac{z^{52}}{4}+\frac{z^{60}}{2}+\cdots,\\
	f(z)&=\frac{1}{z}\frac{F(z)-1}{F(z)+1}=\frac{z^{3}}{2}-\frac{z^{7}}{4}+\frac{3z^{11}}{8}+\frac{3z^{15}}{16}-\cdots.
\end{align*}
Here we introduce $G(z)$ and $g(z)$ satisfying the following relations respectively, $F(z)=G(z^4)$ and $f(z)=z^3g(z^4)$. Thus we have
\begin{align}\label{caraG}
	G(z)=1+z+\frac{z^{3}}{2}+z^{4}+\cdots,\quad g(z)=\frac{z}{2}-\frac{z^{2}}{4}+\frac{3z^{3}}{8}+\frac{3z^{4}}{16}-\cdots.
\end{align}
Then, the non-zero Verblunsky parameters $\xi_{k}$ can be obtained by using the following algorithm.
\begin{align}\label{schurg}
	g_{k+1}(z)=\frac{1}{z}\frac{g_{k}(z)-g_{k}(0)}{1-\overline{g_{k}(0)}g_{k}(z)},\quad \xi_{k}=g_{k}(0)\quad (k\geq 1),
\end{align}
where $g_{1}(z)=g(z)$. By the relationship between $f(z)$ and $g(z)$, we see that $\xi_{k}=\alpha_{4k-1}\quad (k\geq 1)$. Note that $\alpha_{n}=0$ for $n\neq 4k-1$. Thanks to the method and the Verblunsky parameter given in Gr\"unbaum and Vel{\'a}zquez \cite{Grunbaum2012}, we will compute return probability for the Riesz walk in the next section.

\section{Return probability}\label{returnsec}
In this section, we calculate the return probability of the Riesz walk starting from the origin. First, we present our main result.
\vspace*{12pt}
\begin{thm}\label{returnthm}
	For the Riesz walk on $\ZMP$ with an initial state  $\Psi_0=\left[\alpha,\beta\right]^{\mathrm{T}}\delta_0\,(|\alpha|^2+|\beta|^2=1)$ at the origin, the return probability of the origin at time $t$ is given by
	\begin{align*}
		\mu_t(0)=\left\{\begin{array}{ll}
			\frac{|\alpha|^2}{4^p},&t=4^{k_1}\pm 4^{k_2}\pm \cdots\pm 4^{k_p}-1,\\
			\frac{1}{4^p},&t=4^{k_1}\pm 4^{k_2}\pm \cdots\pm 4^{k_p},\\
			\frac{|\beta|^2}{4^p},&t=4^{k_1}\pm 4^{k_2}\pm \cdots\pm 4^{k_p}+1,\\
			1,&t=0,\\
			|\beta|^2,&t=1,\\
			0,&otherwise,
		\end{array}\right.
	\end{align*}
	where $k_1>k_2>\cdots>k_p\geq 1$ and $\delta_m(x)=1\,(x=m),\,=0\,(x\neq m)$.
\end{thm}
\vspace*{12pt}
We should remark that $\mu_t(0)=1/4$ for $t=4^k(k=1,2,\ldots)$. We call ``localization occurs'' if there exists $x\in\ZMP$ such that $\limsup_{t\rightarrow\infty}\mu_t(x)>0$. In our definition, localization occurs for the Riesz walk, since $\limsup_{t\rightarrow\infty}\mu_t(0)>0$.\par
In order to prove Theorem \ref{returnthm}, we explain the following properties $(1)$ and $(2)$ on the evolution of the Riesz walk. After that, we consider an initial state $\Psi_0=\left[1,0\right]^{\mathrm{T}}$. Moreover, we extend the initial state to general form $\Psi_0=\left[\alpha,\beta\right]^{\mathrm{T}}$ at the end of the proof of Theorem \ref{returnthm}.
\begin{enumerate}
	\renewcommand{\labelenumi}{(\arabic{enumi})}
	\item Parity of amplitude: At even times, the amplitude exists only at $\{\Psi^{L}(0),\Psi^{R}(1),\Psi^{L}(2),\linebreak\Psi^{R}(3),\ldots \}$. At odd times, the amplitude exists only at $\{\Psi^{R}(0),\Psi^{L}(1),\Psi^{R}(2),\Psi^{L}(3),\ldots\}$. Furthermore, this gives us a parity of measure for each time. At even times, measures at each location are expressed as $\mu_{2n}(2x)=|\Psi_{2n}^{L}(2x)|^2$ and $\mu_{2n}(2x+1)=|\Psi_{2n}^{R}(2x+1)|^2$. At odd times, measures at each location are expressed as $\mu_{2n+1}(2x)=|\Psi_{2n+1}^{R}(2x)|^2$ and $\mu_{2n+1}(2x+1)=|\Psi_{2n+1}^{L}(2x+1)|^2$.
	\item Evolution: In the evolution from odd time to even time, this can be thought of as a shift operator like $\Psi_{2n+1}^{L}(x)=\Psi_{2n}^{L}(x-1)$ and $\Psi_{2n+1}^{R}(x)=\Psi_{2n}^{R}(x+1)\,(x\geq0)$. In the evolution from even time to odd time, this can be thought of as a coin operator like 
	\begin{align*}
		\begin{bmatrix}
			\Psi_{2n}^{R}(2x-1)\\\Psi_{2n}^{L}(2x)
		\end{bmatrix}=C_x\begin{bmatrix}
			\Psi_{2n-1}^{L}(2x-1)\\\Psi_{2n-1}^{R}(2x)
		\end{bmatrix},
	\end{align*}
	where 
	\begin{align*}
		C_x=\begin{bmatrix}
			\xi_x&\rho_x\\\rho_x&-\xi_x
		\end{bmatrix}.
	\end{align*}
	Here $\xi_x$ is non-zero Verblunsky parameter and $\rho_x=\sqrt{1-\xi_x^2}$. At the origin, we add $\Psi_t(-1)$ to the entire system and set $\xi_0=-1$ and $\rho_0=0$. Since $\xi_0=-1$, the relationship $\Psi_{2n}^{L}(0)=\Psi_{2n-1}^{R}(0)$ holds at the origin.
\end{enumerate}
Based on the above properties, we introduce another QW. Let $\widetilde{\Psi}_n(x)$ be the amplitude of the QW at location $x$ at time $n$. The relationship between the introduced QW $\widetilde{\Psi}_n(x)$ and the Riesz walk $\Psi_n(x)$ is as follows.
\begin{align*}
	&\widetilde{\Psi}_n(x)=\begin{bmatrix}
		\widetilde{\Psi}_n^{L}(x)\\\widetilde{\Psi}_n^{R}(x)
	\end{bmatrix}=\begin{bmatrix}
		\Psi_{2n+1}^{L}(2x-1)\\\Psi_{2n+1}^{R}(2x)
	\end{bmatrix},\\
	&\widetilde{C}=\sum_x\ket{x}\bra{x}\otimes \widetilde{C_x}=\sum\ket{x}\bra{x}\otimes\begin{bmatrix}
		\xi_x&\rho_x\\\rho_x&-\xi_x
	\end{bmatrix},\\
	&\widetilde{S}=\sum_x \left(\ket{x-1}\bra{x}\otimes\ket{R}\bra{L}+\ket{x+1}\bra{x}\otimes\ket{L}\bra{R}\right),
\end{align*}
where $\widetilde{C}$ is a coin operator, $\widetilde{S}$ is a shift operator, and the evolution of the entire system $\widetilde{U}$ is defined by  $\widetilde{U}=\widetilde{S}\widetilde{C}$. Note that the initial state of the introduced QW is the state at time $1$ of the Riesz walk, that is, $\widetilde{\Psi}_0=[1,0]^{\mathrm{T}}\delta_1$.\par
\begin{figure}[htbp]
	\centering
	\includegraphics[clip,width=0.45\linewidth]{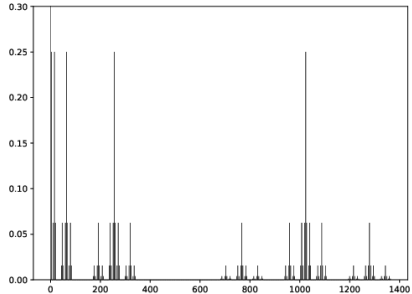}
	\caption{\label{originfig}Probability at the origin from time $0$ to $1365$.}
\end{figure}

We want to calculate the return probability of the Riesz walk. From $\Psi_{2n+1}^{L}(1)=\Psi_{2n}^{L}(0)=\Psi_{2n-1}^{R}(0)=\Psi_{2n-2}^{R}(1)$ by the above mentioned properties, we need to calculate $\Psi_{2n+1}^{L}(1)=\widetilde{\Psi}_{n}^{L}(1)$. To do so, we get the following generating function of  $\widetilde{\Psi}_{n}^{L}(1)$.
\vspace*{12pt}
\begin{lem}\label{returnlem}
	The generating function of $\widetilde{\Psi}_t(x)$ is defined as $\widehat{\widetilde{\Psi}}_x^{M}(z)=\sum_{t=0}^{\infty}\widetilde{\Psi}_t^{M}(x)z^t\,(M\in\{L,R\})$, then $\widehat{\widetilde{\Psi}}_1^{L}(z)$ is given by
	\begin{align*}
		\widehat{\widetilde{\Psi}}_1^{L}(z)=\frac{1+\xi_1\widehat{f}_1^{(+)}(z)}{(1-\xi_1z^2)+(\xi_1-z^2)\widehat{f}_1^{(+)}(z)},
	\end{align*}
	where $\widehat{f}_k^{(+)}$ is the following continued fraction.
	\begin{align*}
		\widehat{f}_x^{(+)}(z)=\frac{z^2(\widehat{f}_{x+1}^{(+)}(z)+\xi_{x+1})}{1+\xi_{x+1}\widehat{f}_{x+1}^{(+)}(z)}=\frac{z^2}{\xi_{x+1}}\left(1-\frac{\rho_{x+1}^2}{1+\xi_{x+1}\widehat{f}_{x+1}^{(+)}(z)}\right).
	\end{align*}
\end{lem}
\vspace*{12pt}
{\bf Proof of Lemma \ref{returnlem}.} From Lemma 3.1 in \cite{KonnoSegawa2013}, we consider $a_x=d_x=\rho_x,\, b_x=-c_x=\xi_x$, and
\begin{align*}
	\widetilde{P}_x=\begin{bmatrix}
		0&0\\b_x&a_x
	\end{bmatrix},\quad \widetilde{Q}_x=\begin{bmatrix}
		d_x&c_x\\0&0
	\end{bmatrix},\quad \widetilde{R}_x=\begin{bmatrix}
		0&0\\d_x&c_x
	\end{bmatrix},\quad \widetilde{S}_x=\begin{bmatrix}
		b_x&c_x\\0&0
	\end{bmatrix}.
\end{align*}
Define $F^{(+)}(x,n)$ (resp. $F^{(-)}(x,n)$) as the weight of all passages starting from location $x$ and returning to $x$ for the first time at time $n$ moving only in $\{y\in\ZM:y\geq x\}$ (resp. $\{y\in\ZM:y\leq x\}$). The generating function $\widehat{F}_x^{(+)}(z)=\sum_{n=2}^{\infty}F^{(+)}(x,n)z^n$ can be expressed as $\widehat{F}_x^{(+)}(z)=\widehat{f}_x^{(+)}(z)\widetilde{R}_x$, where $\widehat{f}_x^{(+)}(z)$ is a complex number. Furthermore, $\Xi^{(+)}(x,n)$ (resp. $\Xi^{(-)}(x,n)$) is defined as the weight of all passages starting from location $x$ and returning to $x$ at time $n$ moving only in $\{y\in\ZM:y\geq x\}$ (resp. $\{y\in\ZM:y\leq x\}$). The generating function $\widehat{\Xi}_x^{(+)}(z)=\sum_{n=0}^{\infty}\Xi^{(+)}(x,n)z^n$ can be expressed as $\widehat{\Xi}_x^{(+)}(z)=I+\widehat{F}_x^{(+)}(z)\Xi_x^{(+)}(z)$, therefore $\widehat{\Xi}_x^{(+)}(z)=\left(I-\widehat{F}_x^{(+)}(z)\right)^{-1}$. In addition, $\widehat{F}_x^{(+)}(z)=z\widetilde{P}_{x+1}\widehat{\Xi}_{x+1}^{(+)}z\widetilde{Q}_x$ also holds, so we have
\begin{align*}
	\widehat{f}_x^{(+)}(z)=\frac{z^2\left(b_{x+1}+\Delta_{x+1}\widehat{f}_{x+1}^{(+)}(z)\right)}{1-c_{x+1}\widehat{f}_{x+1}^{(+)}(z)},\quad\Delta_x=a_xd_x-b_xc_x.
\end{align*}
In a similar way, we get
\begin{align}\label{hatfm}
	\widehat{F}_x^{(-)}(z)=z\widetilde{Q}_{x-1}\widehat{\Xi}_{x-1}^{(-)}z\widetilde{P}_x,\quad \widehat{f}_x^{(-)}(z)=\frac{z^2\left(c_{x-1}+\Delta_{x-1}\widehat{f}_{x-1}^{(-)}(z)\right)}{1-b_{x-1}\widehat{f}_{x-1}^{(-)}(z)}.
\end{align}
Next, $\Xi_y(x,n)$ is defined as the weight of all passages starting from location $y$ arriving at $x$ at time $n$ and $\widehat{\Xi}_{x,y}(z)=\sum_{n=0}^{\infty}\Xi_y(x,n)z^n$ is defined as generating function. From 
$\widehat{\Xi}_{1,1}(z)=I+\left(\widehat{F}_1^{(+)}(z)+\widehat{F}_1^{(-)}(z)\right)\widehat{\Xi}_{1,1}(z)$, we obtain
\begin{align*}
	\widehat{\Xi}_{1,1}(z)=\frac{1}{\gamma_1(z)}\begin{bmatrix}
		1-c_1\widehat{f}_1^{(+)}(z)&a_1\widehat{f}_1^{(-)}(z)\\
		d_1\widehat{f}_1^{(+)}(z)&1-b_1\widehat{f}_1^{(-)}(z)
	\end{bmatrix},
\end{align*}
where $\gamma_1(z)=\mathrm{det}\left(\widehat{F}_1^{(+)}(z)+\widehat{F}_1^{(-)}(z)\right)=1-b_1\widehat{f}_1^{(-)}(z)-c_1\widehat{f}_1^{(+)}(z)-\Delta_1\widehat{f}_1^{(+)}(z)\widehat{f}_1^{(-)}(z)$. Then, inserting $a_0=d_0=0,\quad b_0=-c_0=-1$ into Eq. (\ref{hatfm}), we get $\widehat{f}_1^{(-)}(z)=z^2$ and $\gamma_1(z)=(1-\xi_1z^2)+(\xi_1-z^2)\widehat{f}_1^{(+)}(z)$. Finally, from $\widehat{\widetilde{\Psi}}_1(z)=\widehat{\Xi}_{1,1}(z)[1,0]^{\mathrm{T}}$, we obtain $\widehat{\widetilde{\Psi}}_1^{L}(z)=\left(1+\xi_1\widehat{f}_1^{(+)}(z)\right)/\gamma_1(z)$.$\hspace{\fill}\square$
\vspace*{12pt}\\
By the properties of the Riesz walk, we see that
\begin{align*}
	\widehat{\widetilde{\Psi}}_1^{L}(z^2)=\sum_{t=0}^{\infty}\widetilde{\Psi}_t^{L}(1)z^{2t}
	=\sum_{t=0}^{\infty}\Psi_{2t+1}^{L}(1)z^{2t}
	=\sum_{t=0}^{\infty}\Psi_{2t}^{L}(0)z^{2t}
	=\widehat{\Psi}_0^{L}(z).
\end{align*}
From Lemma \ref{returnlem} and above equation, we have the following generating function of  $\Psi_{n}^{L}(0)$.
\vspace*{12pt}
\begin{pro}\label{returnpro}
	For the Riesz walk on $\ZMP$ with an initial state $\Psi_0=\left[1,0\right]^{\mathrm{T}}\delta_0$, the generating function of amplitude at the origin $\widehat{\Psi}_0^{L}(z)=\sum_{n=0}^{\infty}\Psi_n^{L}(0)z^n$ is given by
	\begin{align*}
		\widehat{\Psi}_0^{L}(z)=\frac{1+\xi_1\widehat{f}_1^{(+)}(z^2)}{(1-\xi_1z^4)+(\xi_1-z^4)\widehat{f}_1^{(+)}(z^2)},
	\end{align*}
	where $\widehat{f}_k^{(+)}$ is the following continued function.
	\begin{align*}
		\widehat{f}_x^{(+)}(z^2)=\frac{z^4(\widehat{f}_{x+1}^{(+)}(z^2)+\xi_{x+1})}{1+\xi_{x+1}\widehat{f}_{x+1}^{(+)}(z^2)}.
	\end{align*}
\end{pro}
\vspace*{12pt}
From now on, we will prove Theorem \ref{returnthm}.
\vspace*{12pt}\\
{\bf Proof of Theorem \ref{returnthm}.} First, the generating function of the Riesz walk is given by
\begin{align}\label{GFofRiesz}
	\widehat{\Psi}_0^{L}(z)&=\frac{1}{\frac{(1-\xi_1z^4)+(\xi_1-z^4)\widehat{f}_1^{(+)}(z^2)}{1+\xi_1\widehat{f}_1^{(+)}(z^2)}}
	=\frac{1}{1-z^4\frac{\xi_1+\widehat{f}_1^{(+)}(z^2)}{1+\xi_1\widehat{f}_1^{(+)}(z^2)}}
	=\frac{1}{1-h_1^{(+)}(z^2)},
\end{align}
where $h_n^{(+)}(z)$ is defined as
\begin{align*}
	h_n^{(+)}(z)=\frac{z^2(\xi_n+\widehat{f}_n^{(+)}(z))}{1+\xi_n\widehat{f}_n^{(+)}(z)}\quad (n\geq 1).
\end{align*}
Noting that
\begin{align*}
	\widehat{f}_n^{(+)}(z^2)=\frac{z^4(\xi_{n+1}+\widehat{f}_{n+1}^{(+)}(z^2))}{1+\xi_{n+1}\widehat{f}_{n+1}^{(+)}(z^2)}=h_{n+1}^{(+)}(z^2)\quad(n\geq 1),
\end{align*}
$h_n(z)$ is the continued functions.
\begin{align*}
	h_n^{(+)}(z^2)=\frac{z^4(\xi_{n}+h_{n+1}^{(+)}(z^2))}{1+\xi_{n}h_{n+1}^{(+)}(z^2)}\quad (n\geq 1).
\end{align*}
On the other hand, it follows from Eqs. (\ref{caraG}) and (\ref{schurg}) that $G(z)=\left(1+zg_1(z)\right)/\left(1-zg_1(z)\right)$ and $g_k(z)=\left(zg_{k+1}(z)+\xi_k\right)/\left(1+\xi_kzg_{k+1}(z)\right)\,(k\geq1)$. Thus,  
\begin{align}\label{cara_halfG}
	\frac{1}{2}\left(G(z^4)+1\right)&=\frac{1}{2}\left(\frac{1+z^4g_1(z^4)}{1-z^4g_1(z^4)}+1\right)=\frac{1}{1-z^4g_1(z^4)}=\frac{1}{1-\widehat{g}_1(z^4)},
\end{align}
where $\widehat{g}_n(z)$ is defined as
\begin{align*}
	\widehat{g}_n(z)=zg_n(z)\quad (n\geq 1).
\end{align*}
Then, $\widehat{g}_{n+1}(z)$ is the following continued function.
\begin{align}\label{hatg}
	\widehat{g}_n(z^4)&=z^4\times\frac{z^4g_{n+1}(z^4)+\xi_n}{1+\xi_nz^4g_{n+1}(z^4)}
	=\frac{z^4\left(\xi_n+\widehat{g}_{n+1}(z^4)\right)}{1+\xi_n\widehat{g}_{n+1}(z^4)}\quad (n\geq1).
\end{align}
Combining Eqs. (\ref{GFofRiesz}) and (\ref{cara_halfG}) with Eq. (\ref{hatg}), we have
\begin{align}
	\widehat{\Psi}_0^{L}(z)&=\frac{1}{2}\left(G(z^4)+1\right)=\frac{1}{2}\left(F(z)+1\right)=\frac{1}{2}\left\{\left(1+2\sum_{n=1}^{\infty}\bar{\mu}_nz^n\right)+1\right\}=\sum_{n=0}^{\infty}\bar{\mu}_nz^n.
\end{align}
Therefore, Eq. (\ref{rieszmoment}) implies that the amplitude at the origin at time $n$ is
\begin{align*}
	\Psi_n^{L}(0)=\left\{ \begin{array}{cl}
		1,&n=0,\\
		1/2^{p},&n=\pm4^{k_1}\pm4^{k_2}\pm\cdots\pm4^{k_p},\quad k_1>k_2>\cdots>k_p\geq1,\\
		0,&otherwise.
	\end{array}\right.
\end{align*}
Furthermore, the properties of the evolution give
\begin{align*}
	\Psi_n^{R}(0)=\Psi_{n+1}^{L}(0)=\left\{ \begin{array}{cl}
		1/2^{p},&n=\pm4^{k_1}\pm4^{k_2}\pm\cdots\pm4^{k_p}-1,\\
		0,&otherwise,
	\end{array}\right.
\end{align*}
where $k_1>k_2>\cdots>k_p\geq1$. Therefore, we see that
\begin{align*}
	\mu_n(0)=\left\{\begin{array}{ll}
		\left|\Psi_n^{L}(0)\right|^2&(n=even),\\
		\left|\Psi_n^{R}(0)\right|^2&(n=odd),
	\end{array}\right.
\end{align*}
so we have the return probability of the Riesz walk with initial state $\Psi_0=[\alpha,0]^{\mathrm{T}}\delta_0$. In the end, considering the Riesz walk with an initial state $\Psi_0=[0,1]^{\mathrm{T}}\delta_0$, the amplitude at time $1$ is $\Psi_1=[1,0]^{\mathrm{T}}\delta_0$. Therefore, from the parity of the amplitude with initial states both $[\alpha,0]^{\mathrm{T}}\delta_0$ and $[0,\beta]^{\mathrm{T}}\delta_0$, we have the desired conclusion.$\hspace{\fill}\square$
\vspace*{12pt}\\
In particular, for an initial state $\Psi_0=[1,0]^{\mathrm{T}}$, we have
\vspace*{12pt}
\begin{cor}\label{returncor}
	For the Riesz walk on $\ZMP$ with an initial state of $\Psi_0=\left[1,0\right]^{\mathrm{T}}\delta_0$, the return probability of the origin at time $t$ is as follows.
	\begin{align*}
		\mu_t(0)=\left\{\begin{array}{ll}
			\frac{1}{4^p},&t=4^{k_1}\pm 4^{k_2}\pm \cdots\pm 4^{k_p}-\delta,\quad k_1>k_2>\cdots>k_p\geq 1,\quad\delta\in\{0,1\},\\
			1,&t=0,\\
			0,&otherwise.
		\end{array}\right.
	\end{align*}
\end{cor}
\vspace*{12pt}
Let $s(k)=\sum_{t=1}^{k}4^t$. In Corollary \ref{returncor}, the return probability from time $4^{k+1}-s(k-1)$ to $4^{k+1}+s(k-1)$ is equal to the return probability from time $4^{k}-s(k-1)$ to $4^{k}+s(k-1)$. Furthermore, quadratic of the return probability from time $4^{k+1}\pm4^k-s(k-1)$ to $4^{k+1}\pm4^k+s(k-1)$ is equal to the return probability from time $4^{k}-s(k-1)$ to $4^{k}+s(k-1)$. Therefore, the return probability has self-similar sets between $4^k-s(k-1)$ and $4^k+s(k-1)$ for each $k$ (see Fig. \ref{originfig}).
\par 
Note that there is another approach of Theorem \ref{returnthm} by using a method in \cite{Cantero2012}. Laurent polynomials $\{\chi_j\}_{j=0}^{\infty}$ are obtained from the Gram-Schmidt orthonormalization of $\{1,z,z^{-1},z^2,\linebreak z^{-2},\ldots\}$. In the case of the Riesz measure, we see that the first moment $\mu_1=0$, so Laurent polynomials are $\chi_0=1,\, \chi_1=z$. Then, the $(j,k)$ element of the evolution $U^{(s)}$ to the power of $n$ is given by
\begin{align*}
	U^{(s)n}_{jk}=\langle\chi_k,z^n\chi_j\rangle.
\end{align*}
Therefore, the elements related to the origin are
\begin{align}\label{EqCGMV}
	&U^{(s)n}_{00}=\langle1,z^n\rangle=\mu_n,\quad U^{(s)n}_{01}=\langle z,z^n\rangle=\mu_{n-1},\nonumber\\
	&U^{(s)n}_{10}=\langle1,z^{n+1}\rangle=\mu_{n+1},\quad U^{(s)n}_{11}=\langle z,z^{n+1}\rangle=\mu_n.
\end{align}
Since the initial state is $\Psi_0=[\alpha,\beta]\delta_0$, the amplitude at the origin at time $n$ is as follows.
\begin{align}\label{EqRP}
	\Psi_n(0)=\begin{bmatrix}
		\alpha\mu_n+\beta\mu_{n-1}\\
		\alpha\mu_{n+1}+\beta\mu_n
	\end{bmatrix}.
\end{align}
Finally, it follows from Eq. (\ref{rieszmoment}) and the above equation that Theorem \ref{returnthm} holds.\par 
For the return probability of the origin, we can obtain the same result by these two approaches. However, we think that it would be difficult to calculate the return probability other than the origin by using the approach in \cite{Cantero2012}, because it is hard to calculate Laurent polynomials $\chi_j$ for large $j$, like our method.
\section{Conjectures on the Riesz walk}\label{conjsec}
In Section \ref{returnsec}, we calculated the probability at the origin only. This section is devoted to conjectures on the evolution of the Riesz walk with initial state $\Psi_0=[1,0]$ based on numerical simulations. Note that we use the values of the Verblunsky parameters in \cite{Grunbaum2012}.\par
From Corollary \ref{returncor}, the probability at the origin at time $4^n$ is $1/4$. On the other hand, the probabilities except the origin can not be calculated. Therefore, we first show the numerical results for the probability distribution at time $4^n$.
\begin{figure}[htbp]
	\centering
	\begin{minipage}{0.45\linewidth}
		\centering
		\includegraphics[clip,width=\linewidth]{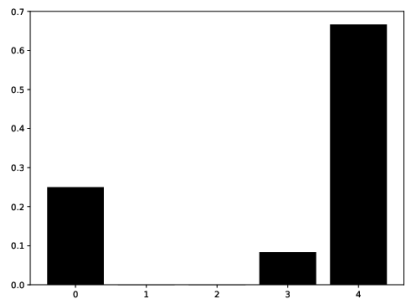}\\
		{\footnotesize (a)}
	\end{minipage}
	\begin{minipage}{0.45\linewidth}
		\centering
		\includegraphics[clip,width=\linewidth]{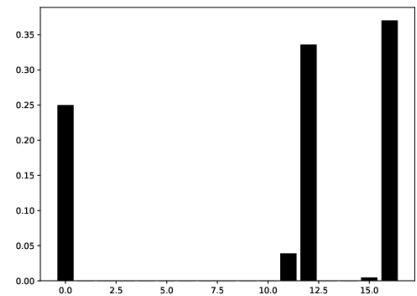}\\
		{\footnotesize (b)}
	\end{minipage}
	\begin{minipage}{0.45\linewidth}
		\centering
		\includegraphics[clip,width=\linewidth]{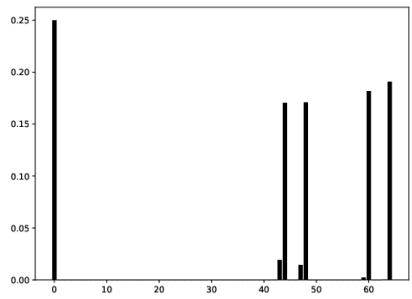}\\
		{\footnotesize (c)}
	\end{minipage}
	\begin{minipage}{0.45\linewidth}
		\centering
		\includegraphics[clip,width=\linewidth]{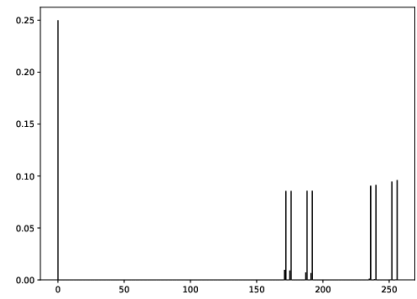}\\
		{\footnotesize (d)}
	\end{minipage}
	\caption{\label{fig_at}Probability distributions of the Riesz walk with initial state $\Psi_0=[1,0]^\mathrm{T}$ for (a) $t=4$, (b) $t=16$, (c) $t=64$, and (d) $t=256$.}
\end{figure}
Let $\nu_t(x)=\mu_t(x-1)+\mu_t(x)$. The probability distribution at time $4$ (Fig. \ref{fig_at} (a)) is $\nu_4(4)=3/4$. The probability distribution at time $16$ (Fig. \ref{fig_at} (b)) is $\nu_{16}(4^2-4)=\nu_{16}(4^2)=3/8$. Furthermore, the probability distribution at time $64$ (Fig. \ref{fig_at} (c)) is
\begin{align*}
	&\nu_{64}(4^3-4^2-4)=0.1895\ldots,\quad\nu_{64}(4^3-4^2)=0.1854\ldots,\\
	&\nu_{64}(4^3-4)=0.1840\ldots,\quad\nu_{64}(4^3)=0.1909\ldots.
\end{align*}
Note that $3/16=0.1875$, which means that the probabilities for each locations are close to $3/16$. Similarly, the probability distribution at time $256$ (Fig. \ref{fig_at} (d)) is that $\nu_{256}(x)$ is close to $3/32$ for $x\in\{4^4-4^3k_1-4^2k_2-4k_3|k_1,k_2,k_3\in\{0,1\}\}$. From these results, we have the following conjecture on the probability distribution at time $4^n$.
\vspace*{12pt}
\renewcommand{\arraystretch}{1.1}
\begin{conj}\label{condist}
	The probability distribution of the Riesz walk on $\ZMP$ at time $4^n$ with initial state of $\Psi_0=[1,0]^{\mathrm{T}}\delta_0$ is in the following. There exists $\varepsilon_{x,n}\in\RM$ with $|\varepsilon_{x,n}|<0.03$ such that
	\begin{align*}
		\mu_{4^n}(x)+\mu_{4^n}(x-1)=\left\{\begin{array}{lcl}
			\frac{1}{4}&,&x=0,\\
			\frac{3}{4}\times\frac{1}{2^{n-1}}\times(1\pm\varepsilon_{x,n})&,&x\in K_n,\\
			0&,&ohterwise,
		\end{array}\right.
	\end{align*}
	where $\RM$ is the set of real numbers and 
	\begin{align*}
		K_n=\left\{4^n-(4^{n-1}k_{1}+4^{n-2}k_{2}+\cdots+4^{1}k_{n-1})\middle|k_{1},k_{2},\ldots,k_{n-1}\in \{0,1\}\right\}.
	\end{align*}
\end{conj}
\renewcommand{\arraystretch}{1}
\vspace*{12pt}
We should remark that numerical simulations suggest $|\varepsilon_{x,n}|<0.03$. The probabilities at time $4^n$ on $K_n$ are not equal to $3/(4\time2^{n-1})$.\par
Next, we consider the evolution by comparing the probability distributions at different times. Comparing the distribution of time $16$ and $64$ (Fig. \ref{fig_at} (b), (c)) in more detail, we find that $\nu_{64}(4^3-4^2-4)+\nu_{64}(4^3-4^2)=\nu_{16}(4^2-4)=3/8$ and $\nu_{64}(4^3-4)+\nu_{64}(4^3)=\nu_{16}(4^2)=3/8$. Comparing the distribution of time $64$ and $256$ (Fig. \ref{fig_at} (c), (d)), we find that $\nu_{256}(x-4)+\nu_{256}(x)=\nu_{64}(x/4)$ for $x\in\{4^4-4^3k_1-4^2k_2|k_1,k_2\in\{0,1\}\}$. In fact, this relationship holds at time not just $4^n$.
\begin{figure}[htbp]
	\centering
	\begin{minipage}{0.45\linewidth}
		\centering
		\includegraphics[clip,width=\linewidth]{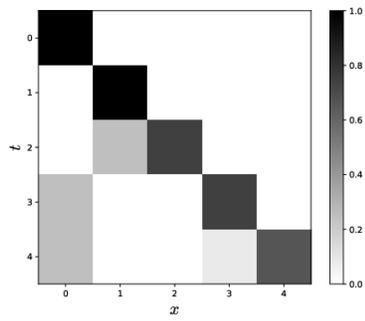}\\{\footnotesize (a)}
	\end{minipage}
	\begin{minipage}{0.45\linewidth}
		\centering
		\includegraphics[clip,width=\linewidth]{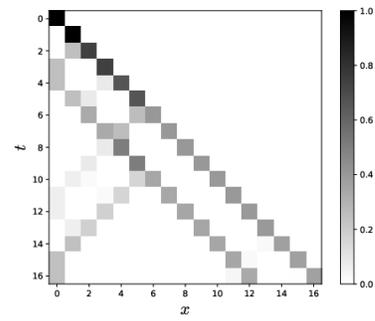}\\{\footnotesize (b)}
	\end{minipage}\\
	\begin{minipage}{0.45\linewidth}
		\centering
		\includegraphics[clip,width=\linewidth]{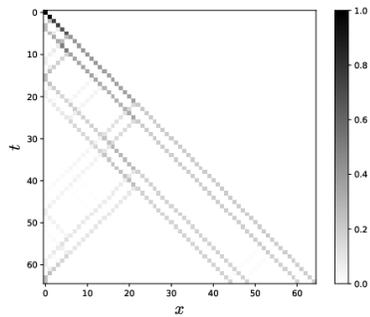}\\{\footnotesize (c)}
	\end{minipage}
	\begin{minipage}{0.45\linewidth}
		\centering
		\includegraphics[clip,width=\linewidth]{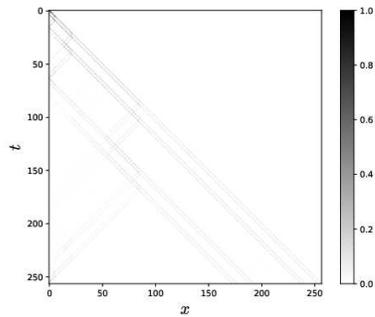}\\{\footnotesize (d)}
	\end{minipage}
	\caption{\label{fig_ts}Probability distribution profile in the plane of position $x$ vs time $t$ until (a) $t=4$, (b) $t=16$,  (c) $t=64$, and (d) $t=256$.}
\end{figure}\par
The time and space spread of probability distribution until $4$ (Fig. \ref{fig_ts} (a)) reproduces the time and space spread of probability distribution until $16$ (Fig. \ref{fig_ts} (b)). Similarly, considering the time and space spreads of probability distribution until times $64$ and $256$ (Fig. \ref{fig_ts} (c), (d)), the time and space spread of the probability distribution is self-similarity until each quadruple time. From this self-similarity, the above relationship at time not just $4^n$ are confirmed by numerical calculation, and the following conjecture about the probability distributions between any even time and its quadruple time is obtained.
\vspace*{12pt}
\begin{conj}\label{consim}
	Let $\ZM_{>}=\{1,2,3,\ldots\}$. For each $n\in\ZM_{>}$, we put $T_0(n)=\{0\}$, and $T_k(n)=((k-1)/n,k/n]\quad(1\leq k\leq n)$. Then, for the measure of the Riesz walk with initial state of $\Psi_0=[1,0]^{\mathrm{T}}\delta_0$,
	\begin{align*}
		P\left(\frac{X_{2t}}{2t}\in T_k(t)\right)=P\left(\frac{X_{8t}}{8t}\in T_k(t)\right),
	\end{align*}
	for any $t\in\ZM_{>}$.
\end{conj}
\vspace*{12pt}\par
We note that Conjecture \ref{consim} means
\begin{align*}
	\mu_{2t}(0)=\mu_{8t}(0),\quad\mu_{2t}(2x-1)+\mu_{2t}(2x)=\sum_{y=8x-7}^{8x}\mu_{8t}(y)\quad(t,x\geq 1).	
\end{align*}\par
Finally, we consider a limit theorem of the Riesz walk. The well-known weak limit theorem of the one-dimensional two-state QW was given by \cite{Konno2002,Konno2005}, whose limit density is an inverse-bell shape. On the other hand, the corresponding limit theorem for QW defined by singular continuous measure is not known. While the authors don't expect such a weak limit theorem, if Conjecture \ref{condist} holds, the following new type of a limit theorem of the Riesz walk might be obtained. Put $\delta_{[a,b]}(x)=1,(x\in[a,b]),=0,(x\notin[a,b])$.
\vspace*{12pt}
\begin{conj}\label{conlimit}
	The limit theorem of the Riesz walk on $\ZMP$ with initial state $\Psi_0=[1,0]^{\mathrm{T}}\delta_0$ in the limit $n\rightarrow \infty$ along time $4^n$ is given by
	\begin{align*}
		\lim_{n\rightarrow\infty}\frac{X_{4^n}}{4^n}=Z,
	\end{align*}
	where $Z$ has the following measure: 
	\begin{align*}
		\frac{1}{4}\delta_0+\mbox{``a self-similar set''}\,\times\delta_{\left[2/3,1\right]}.
	\end{align*}
\end{conj}
\vspace*{12pt}\par
From Theorem \ref{returnthm}, we see that the return probability at the origin at time $4^n$ is $1/4$. From now on, we explain ``a self-similar set'' in Conjecture \ref{conlimit}. First, we consider the Cantor set on closed interval $[0,1]$, which is the well-known self-similar set \cite{Dovgoshev2006}. The Cantor set is constructed by repeatedly removing the middle third of intervals. The Cantor set $C$ is defined by $\cap_{n=0}^{\infty}C_n$, where $C_n$ consists of $2^n$ disjoint closed intervals $C_n^i=[a_n^i,b_n^i]\,(i=1,2,\ldots,2^n)$. The Cantor set can be expressed in other way. Let $R_n$ be a set of right-hand points of intervals $C_n^i\in C_n$, and we have 
\begin{align*}
	R_n=\left\{b_n^i\middle|i=1,2,\ldots,2^n\right\}=\left\{1-2\left(3^{-1}k_1+3^{-2}k_2+\cdots+3^{-n}k_n\right)\middle|k_1,k_2,\ldots,k_n\in\{0,1\}\right\}.
\end{align*}
Then, the Cantor set is $C=\lim_{n\rightarrow\infty}R_n$. Second, as in the case of the Cantor set $C$, we consider the set $D$ that consists of each interval repeatedly divided into four sections and removing the middle two. Similarly, defining a set $M_n$ of right-hand points of interval, we have
\begin{align*}
	M_n=\left\{1-3\left(4^{-1}k_1+4^{-2}k_2+\cdots+4^{-n}k_n\right)\middle|k_1,k_2,\ldots,k_n\in\{0,1\}\right\}.
\end{align*}
Then, $D=\lim_{n\rightarrow\infty}M_n$. Third, when the space is divided by time in Conjecture \ref{condist}, location $K_n$ where the measures are positive is as follows.
\begin{align*}
	\widetilde{K_n}=\left\{1-\left(4^{-1}k_1+4^{-2}k_2+\cdots+4^{-(n-1)}k_{n-1}\right)\middle|k_1,k_2,\ldots,k_{n-1}\in\{0,1\}\right\}.
\end{align*}
Now, we have $(1/3)M_{n-1}+2/3=\widetilde{K_n}$. Therefore, $\widetilde{K_n}$ can be considered as a mapping of $M_{n-1}$ to an interval $[2/3,1]$, and $\lim_{n\rightarrow\infty}\widetilde{K_n}$ is the set of $D$ mapped to interval $[2/3,1]$. Thus, in the limit along $4^n$ of the Riesz walk, the measure lies on ``a self-similar set'' $(1/3)D+2/3 \,(=\lim_{n\rightarrow\infty}\widetilde{K_n})$ in Conjecture \ref{conlimit}.

\section{Generalization of the Riesz walk}\label{SecGeneral}
This section deals with a generalized Riesz walk given by the following singular continuous measures of infinite products of $m$-fold oscillations $(m\geq2)$.
\begin{align*}
	d\mu(z)=\prod_{k=1}^{\infty}\left(1+\cos(m^k\theta)\right)\frac{d\theta}{2\pi}=\prod_{k=1}^{\infty}\left(1+\frac{z^{m^k}+z^{-m^k}}{2}\right)\frac{dz}{2\pi iz}.
\end{align*}
Note that if we take $m=4$, then this becomes the Riesz measure. In order to consider the QW, we need the moments of this measure. One of the important points is that the properties of the moments are different for $m=2$ and $m\geq 3$.\par 
In the case of $m=2$, the moments are given by
\begin{align*}
	\mu_j=\left\{\begin{array}{cc}
		0,&(j=odd),\\
		1,&(j=even).
	\end{array}\right.
\end{align*}
Thus, we obtain the Carath\'eodory function $F(z)=1+2\sum_{j=1}^{\infty}z^{2j}$ and the Schur function $f(z)=z$. Furthermore, the Verblunsky parameters are $\alpha_0=0,\,\alpha_1=1$. Then, the evolution of the QW with $m=2$ becomes 
\begin{align*}
	P_0=\begin{bmatrix}
		0&0\\0&0
	\end{bmatrix},\quad R_0=\begin{bmatrix}
		0&1\\1&0
	\end{bmatrix},\quad Q_1=\begin{bmatrix}
		0&0\\0&0
	\end{bmatrix}.
\end{align*} 
Therefore, even if an initial state at the origin is given, the amplitude stays at the origin for any time. This QW is trivial.\par 
Next, we consider $m\geq3$. In this case, the moments are as follows.
\begin{align}\label{EqMoment-m}
	\mu_{j}=\left\{ \begin{array}{cl}
		1,&j=0,\\
		1/2^{p},&j=\pm m^{k_1}\pm m^{k_2}\pm\cdots\pm m^{k_p},\\
		0,&otherwise,
	\end{array}\right. 
\end{align}
where $k_1>k_2>\cdots>k_p\geq1$. \par 
From the above moments, we obtain the return probability for each QW with $m(\geq 2)$ starting from the origin. For $m\geq 2$, the first and second terms of Laurent polynomials are $\chi_0=1$ and $\chi_1=z$ independent of $m$, because the first moment is $\mu_1=0$. Thus, from Eq. (\ref{EqCGMV}), the amplitude at the origin of a QW $(m\geq 2)$ with an initial state $\Psi_0=[\alpha,\beta]^{\mathrm{T}}\delta_0$ is the same as Eq. (\ref{EqRP}). Therefore, we see that the return probability at the origin can be expressed in terms of moments. In particular, for $m\geq 3$, it follows from the $j$-th moment given by Eq.(\ref{EqMoment-m}) that
\begin{align*}
	\limsup_{t\rightarrow\infty}\mu_n(0)=\frac{1}{4}>0.
\end{align*}
Note that this value $1/4$ does not depend on $m(\geq3)$. We confirm that the localization in our definition occurs at the origin for any QW with $m$. \par 
Furthermore, we expect that similar conjectures, mentioned in Section \ref{conjsec}, hold for general $m\geq 3$.

\section{Conclusion}\label{conclusion}
Non-trivial rigorous results on the Riesz walk are not much known. Combining the CGMV method with the generating function approach, we calculated the return probability of the walk starting from the origin on $\ZMP$. Therefore, it follows from this that the localization occurs in our definition for the Riesz walk. In addition, some conjectures on self-similar properties of the Riesz walk were presented by using numerical simulations. As a future work, it would be fascinating to give proofs of own conjectures and compute the return probability of the Riesz walk starting from any location.

\section*{Acknowledgements}
\noindent
We would like to thank Takashi Komatsu for useful discussion.

\end{document}